\documentclass[%
 reprint,superscriptaddress,
 amsmath,amssymb,
 aps,
]{revtex4-2}

\usepackage{graphicx}
\usepackage{dcolumn}
\usepackage{bm}
\usepackage[colorlinks=false,hidelinks=true]{hyperref}
\usepackage{xcolor}
\usepackage{subfigure}
\usepackage{todonotes}
\usepackage{siunitx}

\usepackage{glossaries}
\newacronym{md}{MD}{molecular dynamics}
\newacronym{ml}{ML}{machine learning}
\newacronym{nn}{NN}{neuronal network}
\newacronym{mlip}{MLIP}{machine-learned interaction potential}
\newacronym{eam}{EAM}{embedded-atom model}
\newacronym{nep}{NEP}{neuro-evolution potential}

\begin{document}

\title{The effect of oxygen on the surface tension of AlTi metallic alloys}

\author{Yuna Kato}
\affiliation{School of Materials and Chemical Technology, Institute of Science Tokyo, Tokyo, Japan}
\affiliation{Institute of Frontier Materials on Earth and in Space, German Aerospace Center, Cologne, Germany}
\author{J\"urgen Brillo}
\affiliation{Institute of Frontier Materials on Earth and in Space, German Aerospace Center, Cologne, Germany}
\author{Thomas Voigtmann}
\affiliation{Institute of Frontier Materials on Earth and in Space, German Aerospace Center, Cologne, Germany}
\affiliation{Institute for Theoretical Physics, Heinrich-Heine-Universit\"at D\"usseldorf, 40225 D\"usseldorf, Germany}
\author{Linnea Heitmeier}
\email{heitmeier@ruc.dk}
\affiliation{Institute of Frontier Materials on Earth and in Space, German Aerospace Center, Cologne, Germany}
\affiliation{Institute for Theoretical Physics, Heinrich-Heine-Universit\"at D\"usseldorf, 40225 D\"usseldorf, Germany}
\affiliation{``Glass and Time'', IMFUFA, Department of Science and Environment, Roskilde University, Postbox 260, DK-4000 Roskilde, Denmark}

\date{\today}

\begin{abstract}
We perform \gls{md} simulations of AlTi metallic alloy melts, with a specific
focus on surface properties and the effect of oxygen. The simulations use a
\gls{mlip} that has recently been tested through experimental data for
the statical and dynamical properties of AlTi melts in the bulk
[Kato \textit{et~al.} (2026)].
We find that the surface-tension response to oxygen depends strongly on the AlTi composition. Element-resolved density profiles show that oxygen enrichment near the interface is accompanied by a suppression of the Al surface segregation observed in the oxygen-free liquid.
The results are compared to experimental values obtained using levitation
techniques in microgravity.
\end{abstract}

\maketitle
\glsresetall

\section{Introduction}

AlTi alloys play a central role in applications ranging from automotive engineering to aerospace and space technologies, where their low density, high strength, and thermal stability are critical \cite{aluminum1984aluminum, davis1993aluminum, reed2008superalloys}. Specifically, surface tension governs key phenomena such as wetting, nucleation, and microstructure evolution during solidification, thereby directly influencing the final material performance \cite{flemings1974solidification, egry2010surface, boinovich2011wetting}.

A major complication in studying metallic melts is that they are never perfectly pure. Even small amounts of impurities, especially oxygen, can significantly alter both surface and bulk properties \cite{keene1993review, brillo2016thermophysical, brillo2019surface, egry2010surface}. Oxygen is known to segregate to surfaces and interfaces, potentially modifying surface tension and inducing structural changes in the near-surface region. Despite its importance, systematic experimental investigations of such effects remain extremely challenging \cite{egry2009surface, ricci1988thermodynamic}. Measurements of surface tension in reactive metallic systems require containerless processing techniques to avoid contamination and heterogeneous nucleation, and often benefit from microgravity conditions to suppress distortions caused by external forces. As a result, available experimental data are scarce and often limited in scope.

In this context, atomistic simulations have emerged as a powerful complementary approach. Classical molecular dynamics (MD) has been widely used to investigate surface properties of liquid metals, including surface tension and interfacial structure. Previous studies—such as those in Refs. \cite{horbach2007self, rozas2016thermophysical, kuhn2013molecular, peng2015structural, heitmeier2026shear} have demonstrated the capability of MD simulations to capture interfacial phenomena. However, many of these investigations rely on embedded atom method (EAM) potentials or simplified model systems, which may limit their predictive accuracy and transferability, particularly for complex, multicomponent alloys with impurities.

Recent advances in \glspl{mlip} have significantly improved the situation \cite{behler2016perspective, sandberg2024homogeneous, mishin2021machine, thiemann2025introduction}. These potentials combine near first-principles accuracy with the computational efficiency required for large-scale MD simulations. In particular, the recently developed neuroevolution potential (NEP) framework has demonstrated high transferability across a wide range of compositions and thermodynamic conditions \cite{liang2025nep89}. This makes it especially suitable for studying alloy systems with varying compositions and impurity levels. For AlTi melts, the NEP has recently been validated against available experimental and computational data \cite{kato2026}, showing good agreement overall, with only minor deviations on the Ti-rich side.

In the present work, we employ this transferable \gls{mlip} to investigate the influence of oxygen impurities on the surface tension of AlTi melts. Using MD simulations in a slab geometry with two approximately non-interacting liquid–vapor interfaces, we compute surface tensions from the anisotropy of the pressure tensor and analyze density profiles across the interface. By systematically varying the oxygen mole fraction within the homogeneous-liquid range, we aim to elucidate how even small impurity levels affect interfacial properties and atomic-scale structure near the surface. This approach enables us to establish a direct link between microscopic arrangements and macroscopic surface behavior, providing insights that are difficult to obtain experimentally.

The remainder of this paper is organized as follows: In section \ref{sec:methods}, we explain the simulation method. After that, we investigate the effect of oxygen on the viscosity and the density of the system in section~\ref{sec:results_bulk}. The surface tension of the system and its behavior at small oxygen concentrations is discussed in section~\ref{sec:results_surfacetension}. After discussing density profiles near the surface in section~\ref{sec:results_density}, we conclude the paper in section~\ref{sec:conclusion}. 

\section{Simulation method}
\label{sec:methods}

We perform \gls{md}-simulations of AlTi, using the \gls{nep} potential developed by \citeauthor{song2024general} \cite{song2024general}. For a detailed description of the simulation method the reader is referred to~\cite{kato2026}. 

\begin{figure}
    \centering
    \includegraphics[width=0.4 \linewidth]{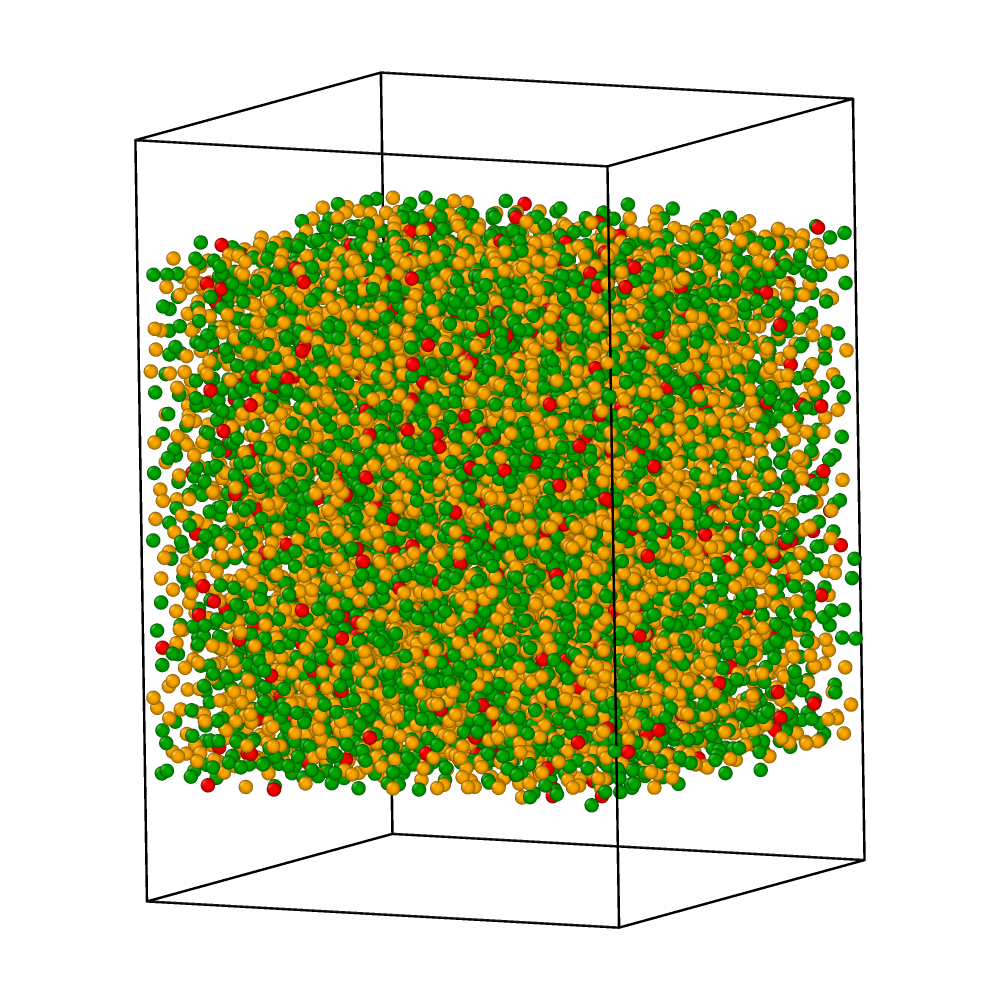}
    \includegraphics[width=0.4 \linewidth]{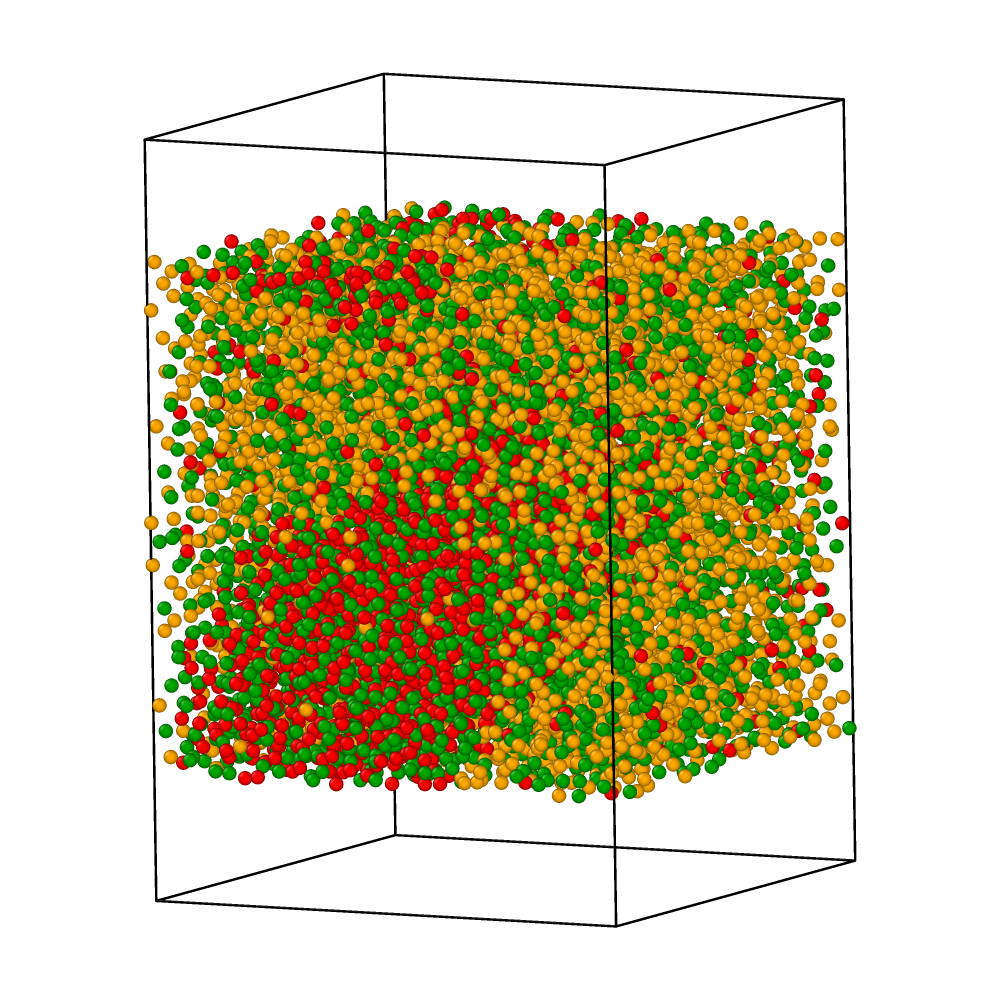}
    \caption{Snapshot of the simulation model used in this study. Yellow and green spheres represent Al and Ti atoms, respectively. Red atoms correspond to the oxygen atoms. Periodic boundary conditions are applied in all Cartesian directions. The configuration on the left-hand side corresponds to $x=0.5$ and $x_O=0.05$. The one on the right-hand side corresponds to $x=0.5$ and $x_O=0.2$. We see a demixing behavior in the case of large oxygen concentrations. Both correspond to a temperature of $T=1950K$.}
    \label{fig:png_01}
\end{figure}

Unless stated otherwise, the number of atoms 
was set to $N=10000$ with periodic boundary conditions in all three spatial 
directions. The equations of motion were integrated with the velocity-Verlet 
algorithm and a time step of $1\ \mathrm{fs}$. 
Each simulation was first equilibrated at $2000\ \mathrm{K}$ for $100\ \mathrm{ps}$, 
and then cooled to the target temperature in the NPT ensemble
(constant particle number, zero external pressure, and linearly decreasing 
temperature). After further equilibration at the target temperature, production 
runs were performed in the NPT ensemble for $200\ \mathrm{ps}$. 
Temperatures $T$ considered in this work were 
$T=1500$, $1600$, $1700$, $1720$, $1800$, $1873$, $1900$, $1950$, $2000$, $2100$, $2200$, $2300$, $2400$,and $2500\ \mathrm{K}$. 
Compositions are specified by two different numbers: We use $x={N_{Al}}/({N_{Ti}+N_{Al})}$ as the ratio between the number $N_{Al}$ of Al-atoms and the total number of metallic atoms, i.e. the sum of $N_{Al}$ and $N_{Ti}$ as the number of Ti-atoms. Additionally, compositions are specified by the oxygen mole fraction $x_O={N_O}/{N_{\text{tot}}}$, specifying the ratio between the number of oxygen atoms and the total number of particles. It follows that the concentration of Al is given by 
\begin{align}
    x_{\text{Al}}=(1-x_O)\cdot x
\end{align}

In this paper, the values of $x$ are given by $(0, 0.5, 1)$, and $x_O$ ranges from $0$ to $0.2$. 



To evaluate the surface tension, after equilibration the simulation box was extended 
along the $z$-direction up to $L_z = 200\ \mathrm{\AA}$, thereby 
creating an interface between the liquid and vacuum regions.
The two-phase system was then equilibrated for another 50 ps.
A sketch of the simulation setup is shown in figure \ref{fig:png_01}. We checked that the resulting surface was stable by investigating the density profiles. 
The surface tension was then calculated from the pressure anisotropy 
\cite{gonzalez2005molecular, heitmeier2026shear, omori2024molecular} as
\begin{align}
    \hat{\sigma} = \frac{1}{2} \int_{-\infty}^{\infty} dz \, \big[ p_n(z) - p_t(z) \big],
\end{align}
where $p_n$ is the pressure component normal to the free surface 
and $p_t$ is the tangential component, and the factor $1/2$ accounts for the
fact that the simulation box contains two separate interfaces.

The surface tension was calculated both for different temperatures and different oxygen mole fractions. 

For the analysis of the interfacial composition, element-resolved density profiles were calculated along the direction normal to the liquid--vacuum interface using a bin width of $0.20$~\AA. The outer boundary of the surface region was defined as the position at which the total atomic density reached 5\% of the bulk-like density when moving from the vacuum towards the liquid. For mixed AlTi compositions, the inner boundary of the surface region was defined as the position at which the local Al fraction among the metallic species returned to the nominal Al fraction of the system. For pure Al and pure Ti, for which such a composition-based boundary cannot be defined, the surface region was taken as the density-transition region between 5\% and 95\% of the bulk-like density. The bulk composition was determined from a 10~\AA-wide region centered at the middle of the liquid slab. The two liquid--vacuum interfaces were analyzed separately, and the resulting surface compositions were averaged.

\section{Results}

\subsection{Bulk structure}
\label{sec:results_bulk}

We begin with a brief discussion of some structural properties, in order to
understand the qualitative effect of oxygen addition in AlTi melts.

    
    

\begin{figure}

\includegraphics[width=\linewidth]{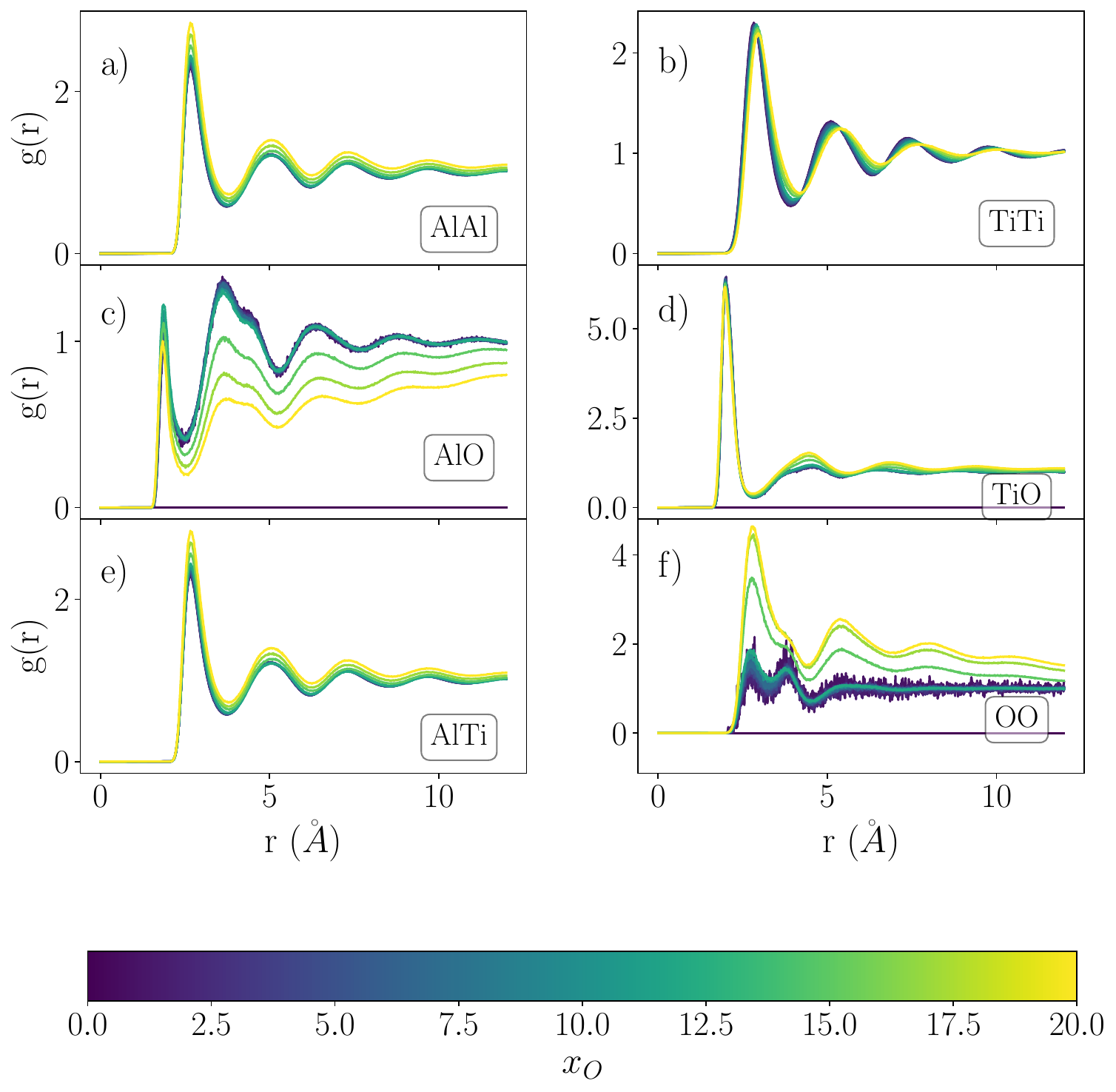}
    \caption{Partial radial distribution functions $g(r)$ at fixed temperature $T=1950 K$ at $x=0.50$. Color codes the O concentration from dark ($x_\text{O}=0\%$) to bright ($x_\text{O}=20\%$).
    \label{fig:gr_o1}
    }
\end{figure}


Figure \ref{fig:gr_o1} shows the partial radial distribution functions as a function of oxygen concentration. While the peak positions of AlAl and AlTi remain nearly unchanged with increasing oxygen concentration, the TiTi peak position clearly shifts. Specifically, an increase of 5\% oxygen concentration results in a 0.8\% elongation of the Ti-Ti distance, and at 10\% oxygen concentration the elongation reaches 2.3\%. This sensitivity indicates that oxygen incorporation preferentially affects the local environment around Ti atoms. The pronounced Ti--O correlation indicates a strong preferential association between Ti and O. The formation of this Ti--O chemical short-range order is accompanied by a shift of the first Ti--Ti peak toward larger distances. Overall, these results show that the oxygen-induced change in bulk chemical short-range order is primarily associated with the local Ti environment.

Additionally, the pair correlation functions of Al-O and O-O indicate a demixing behavior at large oxygen concentrations ($x_\text{O}\gtrsim0.15$ at the specific temperature investigated). This can also be seen in the corresponding simulation snapshot in figure~\ref{fig:png_01}. Here, one can see that in the case of the oxygen concentrations larger than 0.15, the system builds a region with a high amount of Ti and O atoms. This evidence is supported by common Titanium-Oxygen phase diagrams, such as the one of \cite{Gale2004}, which report a liquidus temperature of 2150 K for $x_O=0.2$. This temperature is clearly larger than the 1950 K for which the calculations are done. Thus, the system is not expected to remain within the single-liquid phase region under these conditions. Consistently, the simulation shows the coexistence of a liquid region with an oxygen mole fraction close to 0.2 and a small Ti--O-rich, solid-like domain. 


\begin{figure}
    \centering
    \includegraphics[width=\linewidth]{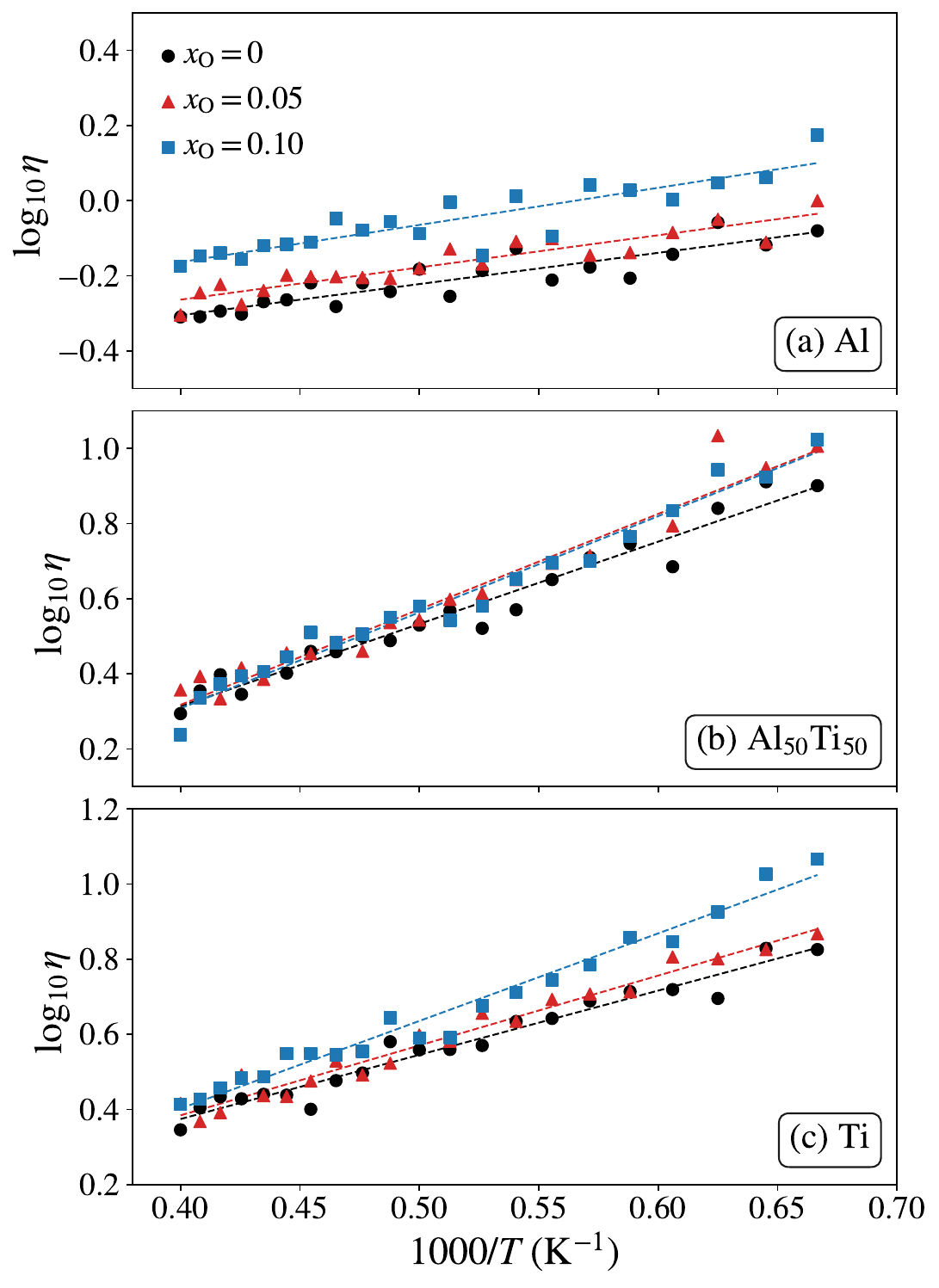}
    \caption{ Logarithm of the viscosity, $\log_{10}\eta$, as a function of inverse temperature for (a) pure Al, (b) equimolar Al$_{50}$Ti$_{50}$, and (c) pure Ti at oxygen mole fractions $x_{\mathrm O}=0$, $0.05$, and $0.10$. Black circles, red triangles, and blue squares represent $x_{\mathrm O}=0$, $0.05$, and $0.10$, respectively. The dashed lines are linear fits to the calculated data. }
    \label{fig:visco_T}
\end{figure}

We investigate the influence of oxygen on the bulk viscosity in Fig.~\ref{fig:visco_T}, where we show the temperature dependence of the viscosity for pure Al, equimolar AlTi, and pure Ti at oxygen mole fractions of $x_{\mathrm O}=0$, 0.05, and 0.10.

Over the investigated temperature range, the temperature dependence of the viscosity can be approximated by an Arrhenius-type relation,
\begin{equation} 
\eta(T) = \eta_0 \exp\left( \frac{E_{\eta}}{RT} \right), 
\label{eq:arrhenius_viscosity} 
\end{equation}
where $\eta_0$ is the pre-exponential factor and $E_{\eta}$ is the apparent activation energy for viscous flow. Taking the common logarithm gives
\begin{equation} 
\log_{10}\eta = \log_{10}\eta_0 + \frac{E_{\eta}}{2.303R} \frac{1}{T}. \label{eq:arrhenius_viscosity_linear} 
\end{equation}
The introduction of oxygen generally increases the viscosity, although the magnitude and temperature dependence of this change depend strongly on the AlTi composition. For pure Al, the slope remains nearly unchanged between $x_{\mathrm O}=0$ and 0.05, with apparent activation energies of 15.9 and 16.4~kJ\,mol$^{-1}$, respectively. At the higher oxygen concentration of $x_{\mathrm O}=0.10$, the apparent activation energy increases to 18.9~kJ\,mol$^{-1}$. However, AlO aggregation becomes apparent in this concentration range. Therefore, the corresponding viscosity cannot be interpreted solely as a property of a homogeneous liquid.
For equimolar AlTi, the apparent activation energy increases from 42.0~kJ\,mol$^{-1}$ in the oxygen-free system to 48.7 and 49.0~kJ\,mol$^{-1}$ at $x_{\mathrm O}=0.05$ and 0.10, respectively. For pure Ti, a progressive increase in the apparent activation energy from 32.8 to 35.6 and 44.7~kJ\,mol$^{-1}$ is observed at $x_{\mathrm O}=0$, 0.05, and 0.10, respectively. Within the homogeneous-liquid range, the oxygen-induced change in the temperature dependence of the viscosity is therefore more pronounced in the Ti-containing liquids than in pure Al.

These observations are consistent with the composition-dependent changes in the local structure discussed above. While the AlAl and AlTi peak positions remain nearly unchanged with increasing oxygen concentration, the TiTi peak position shifts toward larger distances, accompanied by a pronounced TiO correlation. From an atomistic perspective, viscous flow requires continuous rearrangement of the local atomic environment. The oxygen-induced changes in chemical short-range order are therefore reflected in both the magnitude and temperature dependence of the viscosity, although the response depends strongly on the AlTi composition.
The viscosity results thus show that the oxygen-induced changes in local structure are accompanied by composition-dependent changes in bulk transport, with distinct responses observed for pure Al and the Ti-containing liquids.




\subsection{Surface tension}
\label{sec:results_surfacetension}
\begin{figure}
    \centering
    \includegraphics[width=\linewidth]{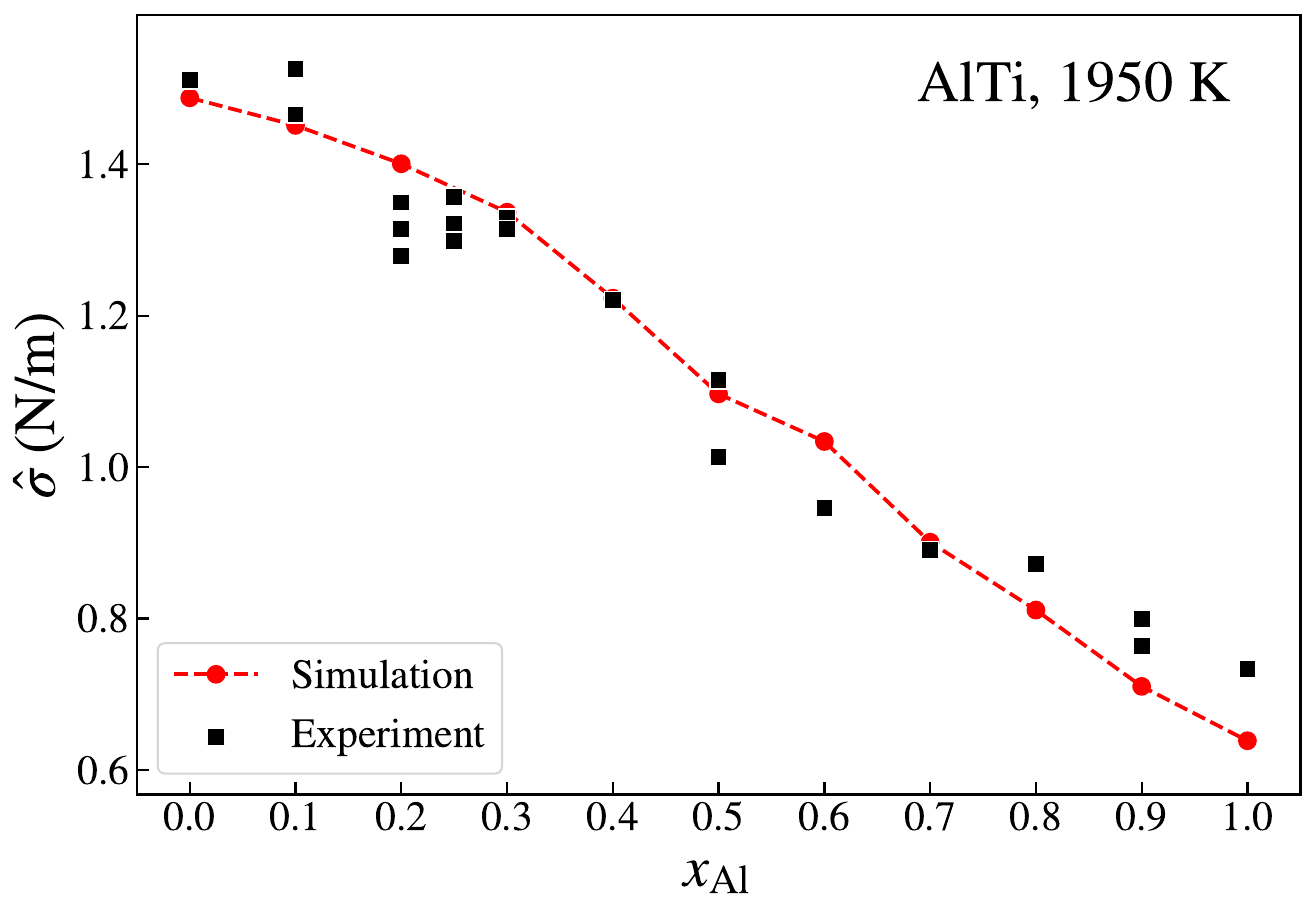}
    \caption{Surface tension against concentration for a system without oxygen at different temperatures. Calculated data are shown as dots and experimental data (1950 K) as squares. Experimental data is taken from \cite{wessing2017density}. 
    }
    \label{fig:st_AlTi_exp}
\end{figure}

We now turn to investigate the effect of oxygen on a free surface. Experimental data, measured containerless by electromagnetic levitation \cite{MohrReviewIJT2024} is available for the nominally pure, i.e. oxygen-free 
AlTi binary alloy \cite{wessing2017density} as well as for the liquid Ti+oxygen system \cite{brillo2019surface,brillowessing2020TiO}. In the latter case, solutions of Ti with defined oxygen mole fractions are produced by adding $\mathrm{TiO_2}$ powder to the Ti-melt and heating the mixture to a temperature of 2200 K, so that all $\mathrm{TiO_2}$ is dissolved / decomposed in the solution.   

For the case without oxygen, a comparison to experiments in Fig.~\ref{fig:st_AlTi_exp} shows that simulation and experiment agree very well, underlining the accuracy of the \gls{mlip} we are using. The accuracy of the interatomic potential is discussed more detailed in Ref.~\cite{kato2026}. 
Regarding the interpretation of the deviations, it should be noted that the simulations were performed in the absence of oxygen, whereas in the experiments the presence of small amounts of oxygen is unavoidable. Nevertheless, deviations coming from oxygen concentrations should be monotonous (as discussed later), but the deviations we see here are not; we thus attribute them to the experimental uncertainty which is specified for the applied oscillating drop technique \cite{MohrReviewIJT2024} as $\Delta \sigma / \sigma \approx \pm 5 \%$.

\begin{figure}
    \centering
    \includegraphics[width=\linewidth]{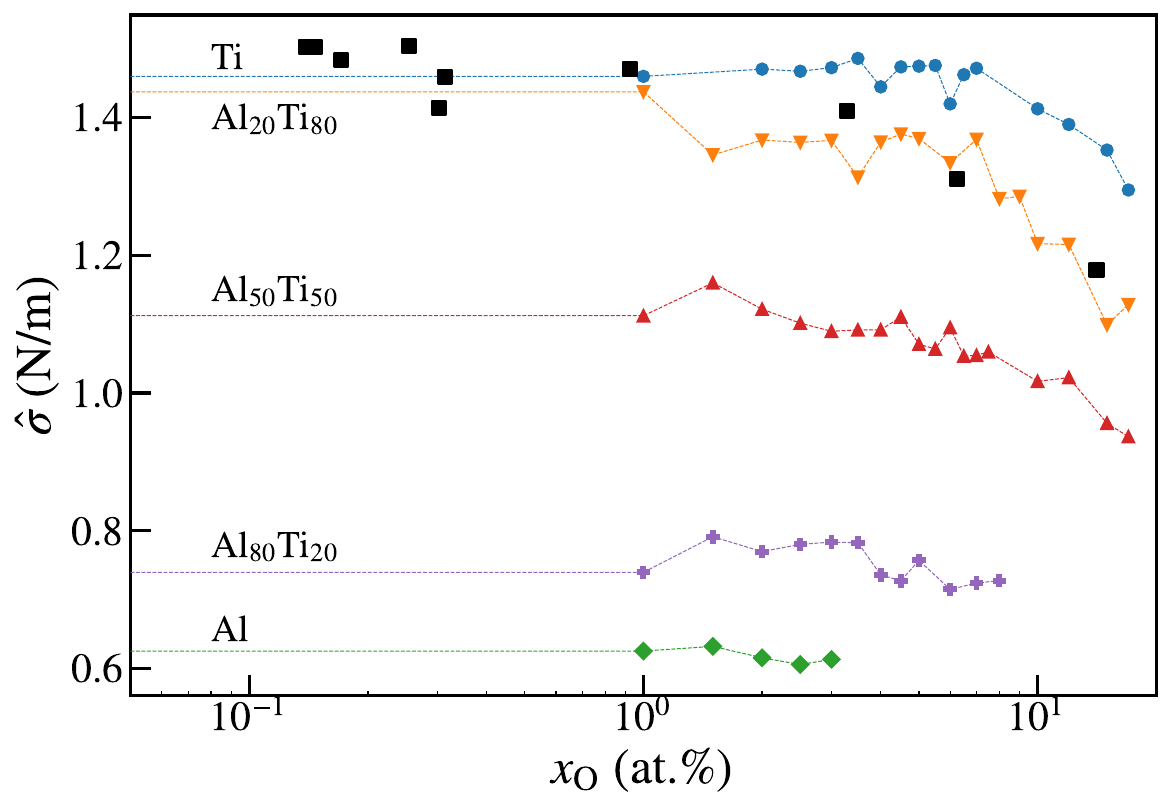}
    \caption{ Surface tension as a function of oxygen mole fraction at $T=1950$~K for liquid AlTiO systems with $x_{\mathrm{Al}}=0$, 0.20, 0.50, 0.80, and 1.00. For each AlTi composition, only the oxygen-concentration range over which a homogeneous liquid state is maintained is shown. The experimental data (black solid squared symbols) for pure liquid Ti are included for comparison and are taken from Ref.~\cite{wessing2018thermophysical}.}
    \label{fig:st_oxygenmix}
\end{figure}

Fig.~\ref{fig:st_oxygenmix} shows the calculated surface tension as a function of oxygen concentration, $x_{\mathrm O}$, for liquid AlTi systems with Al mole fractions $x_{\mathrm{Al}}= 0$, 0.20, 0.50, 0.80, and 1.00. The oxygen concentration is displayed on a logarithmic scale. For pure Ti, the available experimental data\cite{wessing2018thermophysical} are also included for comparison.
The response of the surface tension to oxygen concentration depends strongly on the AlTi composition.

For the Al-rich compositions investigated here, corresponding to $x_{\mathrm{Al}}\geq 0.80$, the surface tension exhibits a comparatively weak dependence on oxygen concentration within the concentration range accessible as a homogeneous liquid. Although some variations are observed for pure Al and Al$_{80}$Ti$_{20}$, these systems do not exhibit the clear and systematic reduction in surface tension observed for the intermediate and Ti-rich compositions.
The oxygen-concentration ranges shown in Fig.~\ref{fig:st_oxygenmix} are not identical for all AlTi compositions. For each composition, only the concentration range over which a homogeneous liquid state was maintained is included. At higher oxygen concentrations on the Al-rich side, pronounced AlO aggregation and the formation of amorphous AlO-rich regions were observed. These oxygen-rich states were therefore excluded from the surface-tension analysis. This behavior suggests that, under the present temperature and composition conditions, the tendency toward AlO aggregation limits the homogeneous-liquid region on the Al-rich side.

In contrast, for the intermediate and Ti-rich compositions investigated here, corresponding to $x_{\mathrm{Al}}\leq0.50$, the surface tension decreases systematically with increasing oxygen concentration. This decrease is particularly pronounced for pure Ti and the Ti-rich alloy, and the calculated trend is qualitatively consistent with the experimental results reported for pure liquid Ti.
For these compositions, two oxygen-concentration regimes can be distinguished. In the low-oxygen regime, below an oxygen mole fraction of approximately $x_{\mathrm O}\simeq 0.06$, the surface tensions remain approximately constant and close to the corresponding values of the nominally oxygen-free systems shown in Fig.~\ref{fig:st_AlTi_exp}. In this regime, the effect of oxygen on the surface tension remains limited.
Above approximately $x_{\mathrm O}\simeq 0.06$, the surface tensions of the intermediate and Ti-rich systems become increasingly sensitive to oxygen and decrease approximately linearly with $\log(x_{\mathrm O})$ \cite{kaptay2015Langmuir,langmuir1917,langmuir1918, szyskowski1908,belton1945}. The same qualitative behavior is observed experimentally for pure Ti \cite{brillowessing2019TiO,brillowessing2020TiO}.

These results demonstrate that the influence of oxygen on surface tension depends not only on oxygen concentration but also on the underlying AlTi composition. In particular, the substantially different responses of the Al-rich and Ti-rich liquids suggest that the distribution of Al, Ti, and O near the free surface plays an important role.

\begin{figure}
    \centering
    \includegraphics[width=\linewidth]{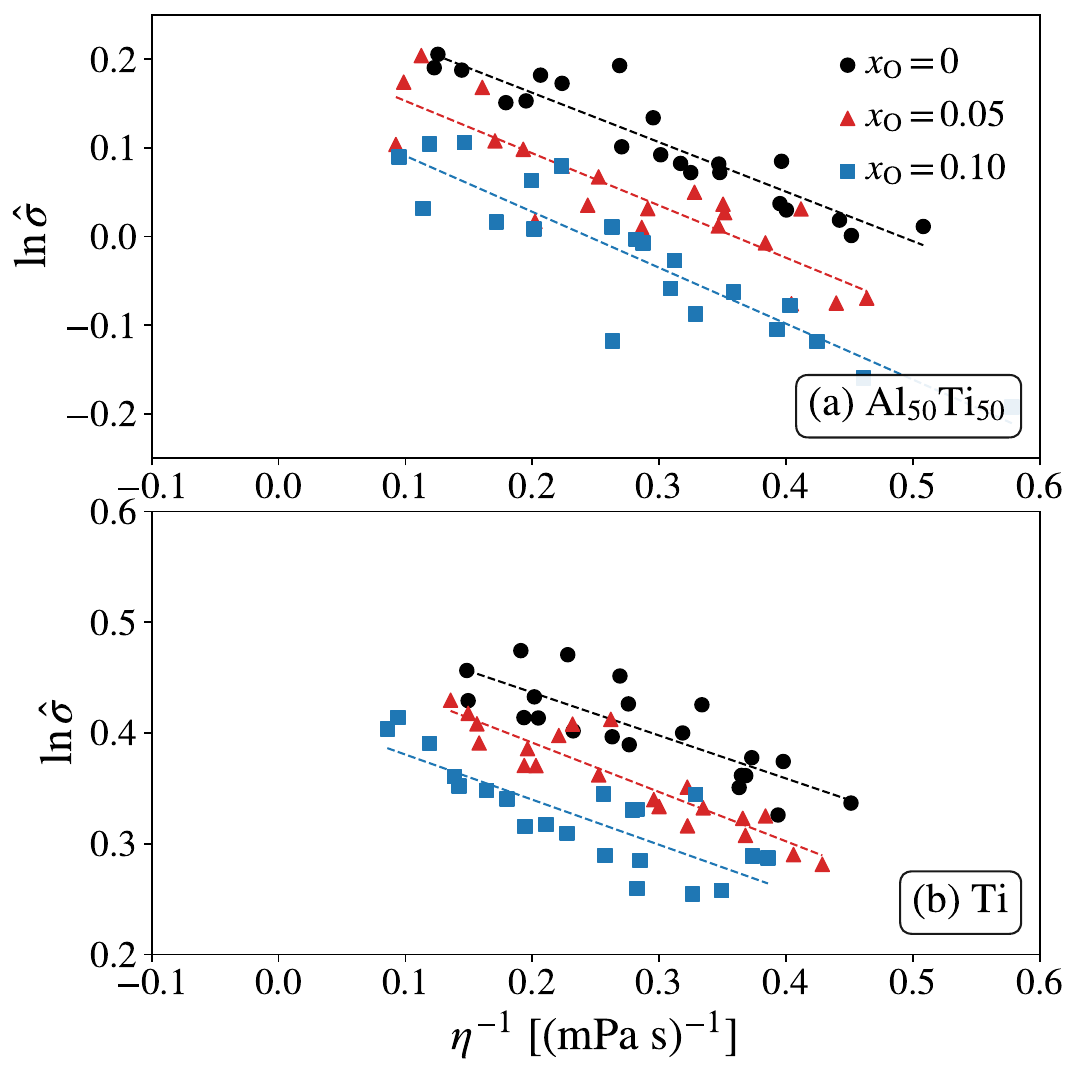}
    \caption{ Relation between surface tension and reciprocal viscosity for (a) equimolar Al$_{50}$Ti$_{50}$ and (b) pure Ti at oxygen mole fractions $x_{\mathrm O}=0$, 0.05, and 0.10. Following \cite{pelofsky1966surface}, we plot $\ln(\hat \sigma)$ against $\eta^{-1}$.}
    \label{fig:surfacetension_visco}
\end{figure}

The relation between surface tension and bulk viscosity is shown in Fig.~\ref{fig:surfacetension_visco}. Following the semi-empirical relations proposed for metallic liquids \cite{egry1992relation,kaptay2005relation,pelofsky1966surface, shimoji1967relation,brillo2016thermophysical}, the relation is expressed in the form 
\begin{equation}
\ln\hat{\sigma} = A\eta^{-1}+B, \label{eq:surface_tension_viscosity} 
\end{equation} 
where $A$ and $B$ are empirical fitting parameters. For each fixed AlTi composition and oxygen concentration, $\ln\hat{\sigma}$ varies approximately linearly with $\eta^{-1}$. The coefficients of determination range from 0.66 to 0.87, indicating that the semi-empirical relation provides a reasonable, although not complete, description of the individual datasets. Pure Al is not included in this comparison because the available oxygen-containing datasets lie outside the homogeneous-liquid range identified above. 

However, the oxygen-containing datasets do not collapse onto the same relation as the corresponding oxygen-free liquids. Within the overlapping range of reciprocal viscosity, the oxygen-containing systems generally exhibit lower values of $\ln\hat{\sigma}$ than the corresponding oxygen-free systems. In addition, the slopes change with oxygen concentration, and the magnitude of this change depends on the underlying AlTi composition. Thus, although a composition-specific empirical correlation between surface tension and reciprocal viscosity is retained within each individual dataset, the relation is not universal across the investigated AlTi-O liquids. Bulk viscosity alone therefore does not provide a unique predictor of surface tension across different oxygen concentrations. The dependence of the relation on both oxygen concentration and AlTi composition suggests that the local distributions of Al, Ti, and O near the free surface must also be considered.



\subsection{Density profiles}
\label{sec:results_density}

\begin{figure}
    \centering
    \includegraphics[width=\linewidth]{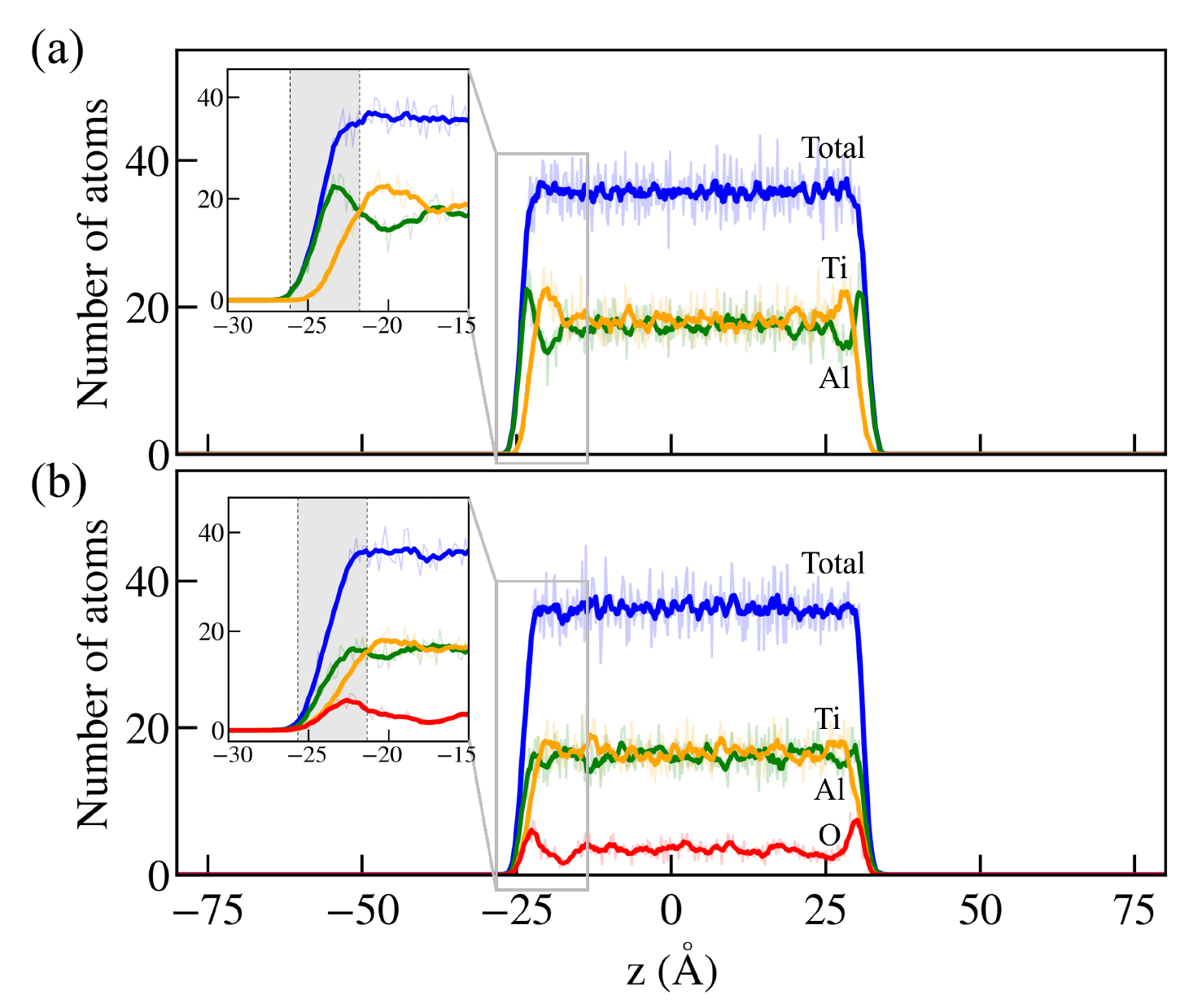}
    \caption{Particle density profile at the free surface: (a) shows the case without oxygen, (b) shows the case with $x_O=0.1$. All plots correspond to a temperature of $1950 K$. The gray shaded regions indicate the outermost surface region used for the subsequent segregation analysis. }
    \label{fig:density_profiles}
\end{figure}

\begin{figure*}[t] 
    \centering 
    \begin{minipage}[t]{0.31\textwidth} 
        \centering \includegraphics[width=\linewidth] {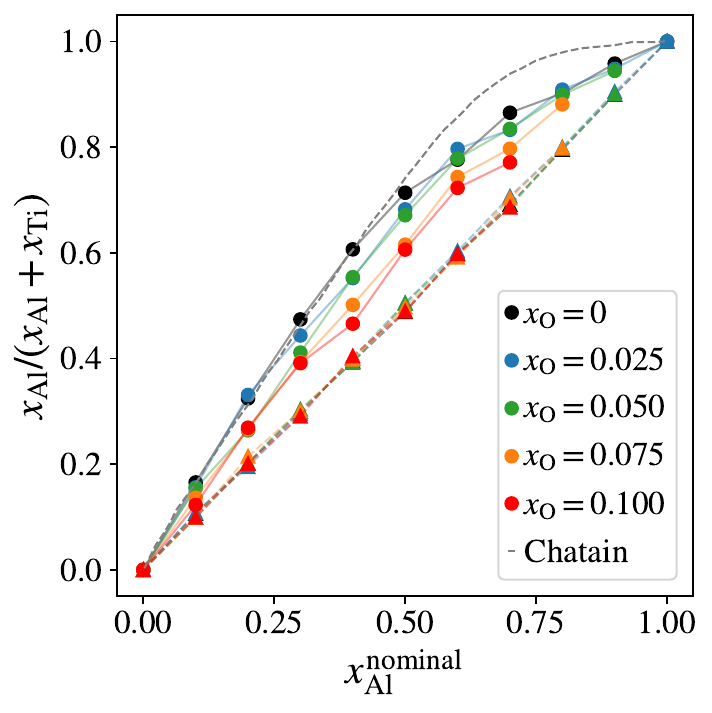} 
    \end{minipage} 
    \hfill 
    \begin{minipage}[t]{0.31\textwidth} 
        \centering \includegraphics[width=\linewidth] {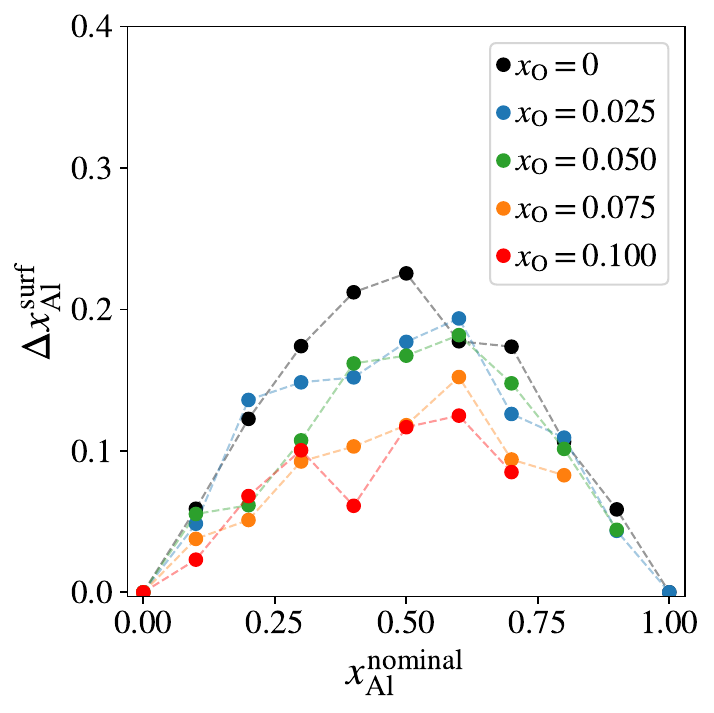}
    \end{minipage} 
    \hfill 
    \begin{minipage}[t]{0.31\textwidth} 
        \centering \includegraphics[width=\linewidth] {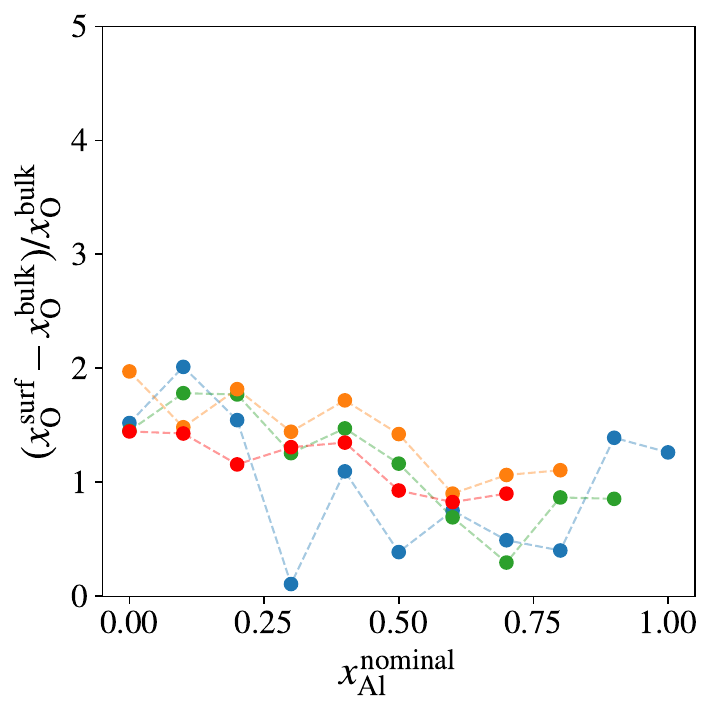} 
    \end{minipage} 
    \caption{ Composition-dependent segregation in the outermost surface region of liquid AlTiO alloys at 1950~K. (a) Al fraction among the metallic species in the outermost surface region. The gray dashed curve shows the first-layer prediction of the Chatain model for the oxygen-free AlTi system.\cite{brillo2016thermophysical} (b) Excess Al fraction in the outermost surface region relative to the bulk-like region, $\Delta x_{\mathrm{Al}}^{\mathrm{surf}} = x_{\mathrm{Al}}^{\mathrm{surf}} - x_{\mathrm{Al}}^{\mathrm{bulk}}$. (c) Relative oxygen enrichment in the outermost surface region, $(x_{\mathrm{O}}^{\mathrm{surf}} -x_{\mathrm{O}}^{\mathrm{bulk}}) /x_{\mathrm{O}}^{\mathrm{bulk}}$. Increasing oxygen content generally reduces Al enrichment near the surface, while relative oxygen enrichment is generally stronger for Ti-rich compositions.  } \label{fig:surface_segregation} 
\end{figure*}

To investigate the microscopic origin of this surface response, we analyze the element-resolved density profiles as a function of distance from the liquid--vapor interface. Comparing the oxygen-free system in Fig.~\ref{fig:surface_segregation}(a) with the oxygen-containing system in Fig.~\ref{fig:surface_segregation}(b), Al is enriched near the surface in the oxygen-free system, whereas oxygen is clearly enriched near the liquid--vapor interface in the oxygen-containing system. In the binary AlTi system, the Al enrichment can be understood from the lower surface tension of liquid Al relative to liquid Ti, which gives rise to a thermodynamic driving force for Al segregation toward the surface \cite{butler1932thermodynamics,brillo2016thermophysical}.

Upon addition of oxygen, however, the surface composition is governed by the competition between all three species. Oxygen preferentially occupies the interfacial region and thereby displaces some of the surface-enriched Al. This redistribution should not be interpreted as implying an increase in surface tension due to the reduced Al surface concentration. Rather, the relevant thermodynamic quantity is the surface free energy of the ternary AlTiO system, for which the energetic stabilization associated with oxygen adsorption at the interface can outweigh the contribution associated with Al segregation. In this sense, oxygen acts as a strongly surface-active component and can substantially lower the surface free energy of the interface.

In addition to this change in surface composition, the oxygen-containing system exhibits a broader interfacial density profile. This broadening indicates that the effect of oxygen is not limited to a simple replacement of surface Al atoms, but is accompanied by a more extended restructuring of the liquid--vapor interface. The microscopic origin of this broadening is discussed further below.

To quantify the composition changes near the surface, we define an outermost surface region from the total-density and local-composition profiles.
The liquid--vapor boundary is defined as the position where the total atomic density reaches 5\% of its bulk-like value.
The inner boundary of the outermost surface region is defined as the position where the local Al fraction returns to the nominal alloy composition.
The gray shaded areas in Fig.~\ref{fig:density_profiles} indicate the outermost surface regions used in the following quantitative segregation analysis.

The thickness of the outermost surface region ranges typically from 3 to 6 \AA, depending mainly on the AlTi composition. This is much smaller than the thickness of the simulated liquid slab, which is approximately 50 \AA. Therefore, the two liquid--vapor interfaces are not expected to influence each other significantly. Moreover, no strong systematic broadening of the outermost surface region with increasing oxygen concentration is observed (Fig. \ref{fig:density_profiles})) The oxygen-induced decrease in surface tension therefore cannot be explained primarily by a change in the interfacial width. We consequently focus on changes in the composition of the outermost surface region.

We first examine whether the profile-based definition captures the surface segregation expected for the oxygen-free binary alloy.
Fig.~\ref{fig:surface_segregation}a compares the Al fraction in the outermost surface region with that in the bulk-like region.
The surface Al fraction exceeds the corresponding bulk-like value over a broad composition range, confirming Al enrichment near the interface.

The oxygen-free simulation is qualitatively consistent with the first-layer prediction of the Chatain model previously applied to liquid AlTi alloys \cite{brillo2016thermophysical}.
The agreement is particularly good at low and intermediate Al concentrations.
Although the thermodynamic surface layer considered in the Chatain model is not identical to the finite-width surface region defined from the present density profiles, the qualitative agreement supports the interpretation of the observed Al enrichment as thermodynamically driven surface segregation.

The Al-rich compositions in Fig.~\ref{fig:surface_segregation}a and Fig.~\ref{fig:surface_segregation}b are retained to illustrate the overall composition dependence. However, the quantitative values at the Al-rich end should be interpreted with caution because the bulk-like reference composition and the inner boundary of the surface region become less well defined.

We next quantify how this intrinsic Al surface enrichment changes upon oxygen addition.
The excess Al fraction in the outermost surface region relative to the bulk-like region is shown in Fig.~\ref{fig:surface_segregation}b.
\begin{equation}
    \Delta x_{\mathrm{Al}}^{\mathrm{surf}} =x_{\mathrm{Al}}^{\mathrm{surf}} - x_{\mathrm{Al}}^{\mathrm{bulk}}
\end{equation}
In the oxygen-free system, $\Delta x_{\mathrm{Al}}^{\mathrm{surf}}$ is positive over a broad composition range and reaches its largest values at intermediate Al concentrations.
With increasing oxygen concentration, the excess Al fraction generally decreases.
Thus, oxygen addition suppresses the intrinsic Al enrichment in the outermost surface region.

The reduction in Al enrichment is accompanied by oxygen segregation toward the liquid--vapor interface.
Fig.~\ref{fig:surface_segregation}c shows the relative oxygen enrichment in the outermost surface region.
\begin{equation}
X_{\mathrm{O}}^{\mathrm{surf}} = \frac{ x_{\mathrm{O}}^{\mathrm{surf}} - x_{\mathrm{O}}^{\mathrm{bulk}} }{ x_{\mathrm{O}}^{\mathrm{bulk}} }
\end{equation}
The relative-enrichment analysis is limited to compositions for which both the surface and bulk-like oxygen concentrations can be determined reliably.
Positive values over most of this composition range demonstrate that oxygen preferentially accumulates near the liquid--vapor interface.

The relative oxygen enrichment is generally stronger for Ti-rich compositions and becomes weaker with increasing Al content. Notably, no preferential oxygen enrichment is observed in pure liquid Al under the investigated conditions.
This composition dependence is consistent with the pronounced short-range Ti-O correlation identified from the partial radial distribution functions in Sec.~\ref{sec:results_bulk} and Fig.~\ref{fig:gr_o1}.
This analysis indicates that oxygen strongly modifies the local environment around Ti atoms. However, the partial radial distribution functions alone do not establish a direct causal relationship between the local Ti-O association and oxygen segregation at the surface.

The combined observations provide a consistent picture of how oxygen modifies the surface segregation behavior.
In the oxygen-free binary AlTi liquid, Al preferentially segregates toward the interface because Al has a lower surface tension than Ti. The resulting deviation from the bulk-like alloy composition lowers the surface free energy. 
In the ternary AlTi-O liquid, oxygen becomes enriched near the interface, while the excess Al concentration decreases. The simultaneous oxygen enrichment, suppression of Al enrichment, and reduction in surface tension suggest that the favorable surface composition shifts from an Al-enriched configuration toward an oxygen-containing configuration. This interpretation is consistent with the lower surface tension observed upon oxygen addition.

\section{Conclusion}
\label{sec:conclusion}
In this study, we investigate the influence of oxygen on both the microscopic structure and surface properties of AlTi metallic alloys. Our findings demonstrate that even small amounts of oxygen induce significant changes in these properties.

In contrast to previous studies that relied primarily on the embedded-atom method (EAM), we employed a \gls{ml} potential trained on ab initio data. Comparison with experimental results reveals excellent qualitative and quantitative agreement, demonstrating the high accuracy and robust applicability of this potential for modeling AlTi systems.

The simulations predict a significant decrease of the surface tension in TiAl
systems with the addition of oxygen. This is in agreement with available
experimental data for liquid Ti. (We are not aware of similar measurements
for TiAl liquid alloys.)
An investigation of spatially resolved density profiles rationalizes this
oxygen-influence on the surface tension as arising from a local displacement
of Al atoms near the surface. Whereas the atomic Ti structure is not
significantly influenced by oxygen concentrations up to $15\%$, AlAl
interactions are very susceptible to the addition of oxygen. We observe
partial phase separation specifically in the AlO substructure around the
largest oxygen concentrations that we studied, indicating that Al and O
displace each other driven by a thermodynamic force.

Our study also has a high experimental relevance: By investigating the influence of oxygen on surface tension, which is small for very small amounts of oxygen and becomes increasingly important as oxygen-concentration of more than $0.1$, we show that the small amounts of oxygen which are inevitable, do not significantly change the measured values. 

Future work will extend this analysis to other Al-based systems, with a particular focus on surface segregation behavior. We aim to determine the composition-dependent conditions under which aluminum segregates to the surface versus the alloying element. Additionally, we intend to investigate whether intrinsic atomic ordering can predict surface dominance in these regions.

Finally, we plan to expand our investigation into interfacial tension. While this study focuses on the liquid--vacuum interface, analyzing the surface tension between metallic alloys and various gases will provide further insights into the fundamental material properties of these liquids.

\section{Data availability}
The data that support the findings of this article are
openly available \cite{zenodooxygen}. 

\begin{acknowledgments}
Y.K. acknowledges financial support from the Dentsu Ikueikai Foundation.
The authors gratefully acknowledge the scientific support and HPC resources
provided by the German Aerospace Center (DLR). The HPC system CARO is
partially funded by ``Ministry of Science and Culture of Lower Saxony'' and
``Federal Ministry for Economic Affairs and Climate Action''.
Part of this work was funded by the Deutsche Forschungsgemeinschaft (DFG,
German Research Foundation) -- project number 505695238, part of the
ANR-DFG project SOLIMAT.
Part of this work was supported by the Independent Research Fund Denmark through grant no. 5281-00148B.
\end{acknowledgments}

\bibliography{sample}

\end{document}